# The bright side of defects in MoS$_2$ and WS$_2$ and a generalizable chemical treatment protocol for defect passivation

Hope M. Bretscher[1], Zhaojun Li[1,2], James Xiao[1], Diana Y. Qiu[3], Sivan Refaely-Abramson[4], Jack Alexander-Webber[5], Arelo O.A. Tanoh[1,6], Ye Fan[5], Géraud Delport[1], Cyan Williams[6,7], Samuel D. Stranks[8], Stephan Hofmann[5], Jeffrey B. Neaton[9,10], Steven G. Louie[9,11], Akshay Rao[1]*

**Affiliations**

[1]Cavendish Laboratory, University of Cambridge, Cambridge, UK.
[2]Molecular and Condensed Matter Physics, Department of Physics and Astronomy, Uppsala University, Uppsala, Sweden.
[3]Department of Mechanical Engineering and Materials Science, Yale University, New Haven CT, USA.
[4]Department of Materials and Interfaces, Weizmann Institute of Science, Rehovot, Israel.
[5]Department of Engineering, University of Cambridge, Cambridge, UK.
[6]Cambridge Graphene Centre, University of Cambridge, Cambridge UK.
[7]Department of Chemistry, University of Cambridge, Cambridge, UK.
[8]Department of Chemical Engineering & Biotechnology, University of Cambridge, Cambridge, UK.
[9]Department of Physics, University of California Berkeley, Berkeley, CA, USA.
[10]Molecular Foundry, Lawrence Berkeley National Laboratory, Berkeley, CA, USA.
[11]Materials Sciences Division, Lawrence Berkeley National Laboratory, Berkeley, CA, USA.
*Correspondence to: ar525@cam.ac.uk

**Abstract:**

Structural defects are widely regarded as detrimental to the optoelectronic properties of monolayer transition metal dichalcogenides, leading to concerted efforts to eliminate defects via improved materials growth or post-growth passivation. Here, using steady-state and ultrafast optical spectroscopy, supported by *ab initio* calculations, we demonstrate that sulfur vacancy defects act as exciton traps. Current chemical treatments do not passivate these sites, leading to decreased mobility and trap-limited photoluminescence. We present a generalizable treatment protocol based on the use of passivating agents such as thiols or sulfides in combination with a Lewis acid to passivate sulfur vacancies in monolayer MoS$_2$ and WS$_2$, increasing photoluminescence up to 275 fold, while maintaining mobilities. Our findings suggest a route for simple and rational defect engineering strategies, where the passivating agent varies the electronic properties, thereby allowing the design of new heterostructures.

**Main Text:**

Single and few-layer semiconducting, two-dimensional transition metal dichalcogenides (2D-TMDCs) have received significant attention in recent years due to their unique optoelectronic properties. Monolayer TMDCs possess a direct bandgap, with optical excitations in the visible range, and very high absorption coefficients [1–3]. Yet studies of their optical and electronic properties and possible device applications have been hindered by the prevalence of defects in these materials, which are proposed to quench photoluminescence (PL) and limit carrier mobilities [4–8].

In the case of Si based electronics improved materials growth was coupled with a systematic understanding of the nature of defects and the development of defect passivation strategies, ultimately leading to extremely small defect densities in electronic devices widely used today. A similar multifaceted strategy is called for with TMDCs. Although there has been advances in materials growth in the past years, our understating of defects and how they degrade performance is still unsatisfactory.

While a number of structural defects have been identified in 2D-TMDCs, chalcogen vacancies are considered to have the lowest formation energy [9–11]. Transmission electron microscopy (TEM) and scanning tunneling microscopy (STM) have been used to identify sulfur vacancies in $MoS_2$ and $WS_2$, occurring at densities of ~$10^{13}$ cm$^{-2}$ [5,11–13]. Recent research combining theory, scanning tunneling spectroscopy (STS) and atomic force microscopy (AFM) techniques, argues that oxygen substitutions at the chalcogen site—which appear nearly identical to chalcogen vacancies in TEM and AFM—are also abundant in TMDCs and are formed both during growth and during exposure to ambient conditions [14–16]. While sulfur vacancies are predicted to result in a state within the electronic bandgap, oxygen substitutions are not expected to possess such a subgap state [10–12,14–20]. The nature of defects remains a topic of hot debate.

Photoluminescence quantum yield (PLQY) can be used as a metric to describe material quality. The PLQY in monolayer $MoS_2$ and $WS_2$ prepared via exfoliation or chemical vapor deposition is very low, often measured below 1 % for $MoS_2$ and only slightly higher for $WS_2$ [1,4,21]. A number of methods have been proposed to increase the PLQY in such samples, including chemical treatments [22–25] and thermal annealing [26,27]. However, the magnitude of the PL increase is often <10x, and no clear consensus has been reached as to how to passivate defects. The most successful chemical treatment is based on the use of the organic superacid bis(trifluoromethane)sulfonimide (TFSI), which was found to increase PL by orders of magnitude [21]. Another study on capacitively gated samples concluded that both the superacid treatment and electrical gating in a capacitive structure increase PL via the same pathway, namely by reducing the high n-doping often found in both exfoliated and grown $MoS_2$ and $WS_2$ monolayers [15,28–31]. These excess charges readily form trions with excitons, for which radiative recombination is much less efficient than for the neutral exciton [20,32,33].

However, these treatments also have their limitations. The PL lifetime increases with TFSI-treatment, which has been shown to be due to the PL still being trap limited [34–36]. This longer PL decay lifetime is not ideal from a device point of view as it opens up the possibility of competing channels of non-radiative decay processes. Carrier mobilities have also been shown to be limited in TFSI treated samples as compared to as exfoliated samples, which further limits the application of such treatment to devices where high carrier mobilities are required [36]. These observations suggest that such treatments, while enhancing PL yield, do not passivate defect sites in the materials completely.

Here, we explore the nature of defect states and chemical passivation methods in TMDC monolayers through steady-state and ultrafast spectroscopy, supported by *ab initio* GW and Bethe Salpeter Equation (GW-BSE) calculations. We experimentally demonstrate that the TFSI superacid treatment leads to the formation of subgap absorption features consistent with theoretically predicted energies for excitons forming from quasiparticle states associated with sulfur vacancies. These shallow subgap states limit PL lifetimes and yields, as well as carrier mobilities. Based on these observations, we develop a new generalized chemical passivation protocol, first treating samples with a passivating agent (PA), such as thiols or sulfides, followed by the TFSI superacid to remove excess electrons (Fig. 1E). This two-step passivation treatment greatly enhances PL, an enhancement of over 275 fold from the brightest spot on a number of untreated samples compared to the brightest spot on treated samples, but also decreases the PL lifetime and improves carrier mobilities relative to TFSI treatment alone, suggesting a passivation of defect sites. The generalizability of this treatment, which can be performed with a number of chemical agents, not only enhances the optoelectronic properties of TMDCs through passivation, but also allows similar chemistries to functionalize these materials and tune their properties.

We begin by investigating the mechanism of the PL enhancement by the superacid TFSI treatment on monolayer $MoS_2$. Samples are mechanically exfoliated onto $Si/SiO_2$ or fused silica substrates using the gold-assisted method [37,38] [details of sample preparation and experimental methods are available in the supplementary materials]. All measurements are performed at room temperature. In line with prior studies, we see a large enhancement in PL when samples are treated with TFSI. In addition, the linewidth narrows and peak-emission energy blueshifts (Fig. 1B, Fig S2) [38]. Since monolayer TMDCs are known to show spot-to-spot variation in the optical response, we measure both the PL lifetime and the distribution of lifetimes over a number of points on multiple flakes, as shown in Fig. 3A. We note that the distribution of average lifetime is large, between 1-20 ns, which is much longer than the lifetime of untreated $MoS_2$, which falls below the instrument response of 100 ps (Fig. S2) [38]. This longer lifetime upon TFSI treatment is consistent with previous results and has been suggested to arise due to a trap mediated exciton recombination process following TFSI treatment [34]. Fig. 1A compares the steady-state optical absorption spectra of TFSI-treated and untreated $MoS_2$. Both samples possess the characteristic A and B exciton peaks at ~1.9 eV and ~2.0 eV respectively. However, in the TFSI treated sample, we find a new subgap absorption feature around 1.7 eV. The emergence of this subgap bright state accompanies the increase in PL yield. To better understand the origin of this state, we compare the experimentally measured absorption to the theoretically computed absorption for a freestanding monolayer of MoS$_2$ with a uniform 2 % concentration of sulfur vacancies calculated within the *ab initio* GW-BSE formalism [38–42]. Consistent with previous calculations on MoSe$_2$ [19], the introduction of sulfur vacancies in monolayer MoS$_2$ gives rise to shallow subgap features ($X_{D2}$) corresponding to excitons arising from transitions between the defect and pristine quasiparticle states. This finding is in good agreement with the subgap features observed in experiment, and in stark contrast to the calculations for oxygen substitution at the defect site, which does not introduce any subgap states in the calculated absorption spectrum (Fig. 1C). This suggests that surprisingly, the TFSI treatment 'opens-up' sulfur vacancy (SV) sites. Fig. 1E gives a schematic of the quasiparticle bands associated with various defect configurations near the K and K' valleys in monolayer MoS$_2$. The sulfur vacancy introduces an occupied defect state close to the valence band edge and an unoccupied defect state deep in the band gap. We speculate that the TFSI treatment activates these defect sites, perhaps by removing substituted oxygen atoms which have been shown not to introduce subgap defect states [14,15,26], and which, according to our calculations, do not introduce subgap features in the optical absorption spectrum either (Fig. 1C). Alternatively, charges from dopants might shift the Fermi level, occupying the subgap defect states and rendering them inaccessible to optical excitations.

To further study the energy-resolved, time evolution of the excited states in MoS$_2$, we performed ultrafast pump-probe measurements. In these measurements, samples are excited with a narrowband, close-to-resonant pump pulse at 1.92 eV, and probed using a broadband white-light pulse [38]. As shown in Fig. 2 A-B, both the untreated and TFSI-treated samples possess positive features around 2.03 eV and 1.88 eV. As the change in transmission is proportional to the change in the density of states, we assign these positive features to the bleach of the A and B excitons. When A and B exciton states are populated, fewer photons are absorbed from the probe pulse, resulting in a positive signal in the differential pump-probe measurement [43]. In the TFSI-treated sample, Fig. 2B, a positive feature in the near-IR appears at the same energy as the state previously observed in the steady-state absorption measurements (Fig. 1A). We attribute this feature to a bleach of the subgap defect state. This defect state is lower in energy and delocalized in momentum space, so excitons could be expected to funnel into these states. Indeed, we observe the A exciton lifetime to be on the order of tens of picoseconds, an order of magnitude shorter than untreated MoS$_2$ (Fig. 2D). Simultaneous with the fast A exciton decay is a growth of the subgap defect bleach, confirming a transfer in population from the band edge A excitons to the defect states. As the excited-state excitons localize in these low-energy states, absorption into those states is reduced, thereby increasing the bleaching signal. The hybridization of states is further corroborated by pump-probe measurements with pump energies resonant with the defect states (Fig S3). While TFSI samples show clear bleach signatures from time zero, untreated samples exhibit negligible signal [38].

Although the defect states absorb light, they do not efficiently emit. When excited above the bandgap, the emission from TFSI-treated samples is dominated by photons near the optical band edge, as shown in Fig. 1B. Surprisingly, with subgap excitation below the band edge, resonant with the defect-state absorption at 1.70 eV, we also observe emission from the band edge at 1.88 eV (Fig. 1D) [38]. This is consistent with previous reports that upconversion from trap states in TFSI treated $MoS_2$ may be responsible for the enhanced PL [34]. Our comparisons of the experiment and theoretical optical absorption spectra suggest that the previously proposed traps are in fact SVs present in the material, which become accessible upon TFSI treatment. Interestingly, based on our observations, upconversion from the defect state to the A exciton state must happen at a faster timescale than the radiative and nonradiative decay of the defect exciton. As discussed in the SI, we estimate, based on the Raman modes of the system, that thermal repopulation of the A exciton is possible on a timescale of hundreds of ps to a nanosecond, giving us a bound on the defect exciton decay rate.

Thus, we find that emission occurs in spite of the presence of sulfur-vacancy-related subgap states. These subgap states are shallow; they trap excitons and prolong the PL lifetime. At room temperature, there is sufficient energy to thermalize to the band edge, where excitons emit. In addition, the lowering of the Fermi level in the presence of the subgap SV sites is detrimental to carrier mobilities, which is why improved charge transport properties are not seen upon TFSI treatment. Thus, despite the increase in PL the TFSI treatment does not serve to actually passivate SV defects in these materials.

With these insights on the role of SVs in untreated and TSFI treated $MoS_2$, we explore other chemical means for passivating defects. Recent theoretical predictions suggest that a thiol bound to the SV in $MoS_2$ would push the site energy above the bandgap, hence effectively passivating the SV [44,45]. Previous studies attempting to use thiol groups to passivate SVs in TMDCs have found them to be ineffective, resulting in reductions in both PL and mobility [46]. We speculate that this is because thiols are a minor n-dopant [47] [48]. When applied to $MoS_2$, though they may passivate defects, they increase the proportion of trions, whose presence limits the quantum yield. We therefore treat samples with a combination of a passivating agent (PA), which we show below can be from a range of different chemicals, and a strong Lewis acid, TFSI (Fig. 4C, Fig. S5).

The PL enhancement obtained by this two-step treatment is on average twice as high as with the TFSI only treatment, as shown in Fig. B and Fig. S2B [38]. The maximum PL spectra observed on the PA+TFSI treated sample is ~275 times brighter than the brightest point on untreated samples (Fig. 1B). Similar to the TFSI treated samples, the peak PL position blueshifts by roughly 30 meV, consistent with a reduction in trions. We map the PL intensity and lifetimes at different points across two flakes, as shown in Fig. 3A. The overall PL intensity is enhanced while the lifetime distribution narrows for each flake (giving the image a bimodal distribution) and lifetime remains under 5 ns (Fig. 3B). Theoretical predictions of the PL lifetime for $MoS_2$ at room temperature are in the 500 ps to a few ns time range [49,50]. Thus, the lower lifetimes obtained via the two-step treatment are more in line with what would be expected from the intrinsic, non-trap limited PL decay. Turning to the steady-state absorption (Fig. 1A), no subgap states are observed for the two-step treatment, in contrast to the TFSI treatment. Similarly, in our pump-probe measurements (Fig. 2C), we observe a greatly reduced subgap bleach, which as discussed above arises from the subgap SV sites. This shows that the two-step treatment greatly passivates these SV sites, while increasing PL yield above that achieved via TFSI only treatment. Further optimization on these treatments should allow for complete passivation of the SV sites, giving both higher PL yield and shorter (more intrinsic) lifetimeThe elimination of trap-limited dynamics is further illustrated by transistor measurements, measuring free carrier mobility. TFSI-treated sample field-effect mobility decreased by two orders of magnitude upon treatment (Fig. 3C). This could be partially due to a decrease in doping, as seen also in the threshold voltage shift (Fig. 3D). However, PA+TFSI treated samples exhibit field-effect mobilities of the same order of magnitude as untreated devices, despite a threshold voltage change comparable to TFSI-treated samples. Therefore, we attribute this decrease in mobility in TFSI-treated $MoS_2$ to the prevalence subgap SV sites, which can be greatly passivated by the two-step treatment developed here, thereby increasing mobilities relative to TFSI treatment alone.

We conclude by discussing the generalizability of the treatment methods presented in this work. First, in Fig. 4A we show that the initial passivation step is achievable using a range of chemicals with sulfur in the -2 oxidation state, such as sulfides and thiols. Second, as shown in Fig. 4B, the PL enhancement is effective in both $MoS_2$ and $WS_2$, showing that SVs can be passivated in a similar manner across different materials.

This work then opens up new avenues for defect engineering of TMDCs, by tuning material properties via a simple, solution-based method and demonstrating that rational passivation strategies can allow for defects to be used as a handle, rather than a hindrance. Along with improvements in properties like PL and mobility, the use of a wide variety of molecules to passivate SV sites opens up the possibility to produce a new library of heterostructures, where the side chains on passivation agents can be chosen for a specific function, such as energy or charge transfer, or to introduce new electronic or magnetic properties.

Fig 1.

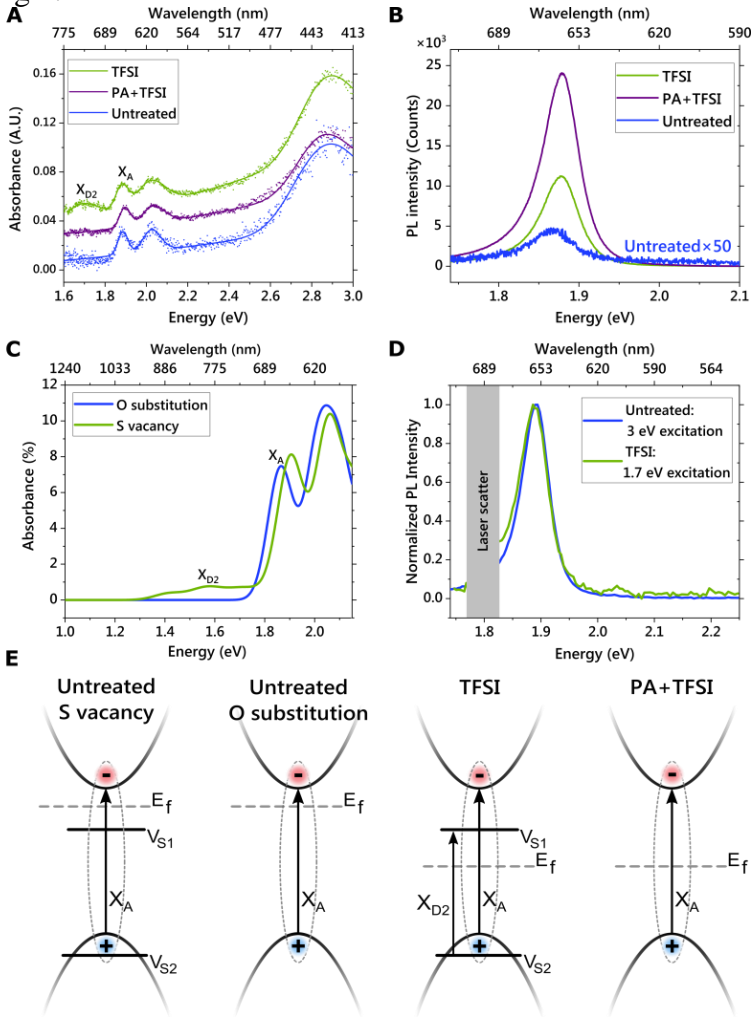

**Figure 1. PL enhancement and subgap state appearance**. (**A**) Absorption of untreated (blue), TFSI treated (green), and PA+TFSI (purple) treated monolayer $MoS_2$. TFSI treated sample possess a subgap absorption peak, $X_{D2}$. (**B**) The PA+TFSI chemical treatment enhances the PL more than TFSI alone. (**C**) Calculated *ab initio* GW-BSE absorbance spectrum of monolayer $MoS_2$ with a 2 % sulfur vacancy (green line) and a 2 % oxygen substitution (blue line) [38], $X_{D2}$ corresponds to excitons arising from transitions between the sulfur vacancy defect and pristine quasiparticle states. (**D**) PL emitted with subgap excitation of TFSI treated $MoS_2$ occurs at the same energy as PL emitted by above-bandgap excitation of untreated $MoS_2$ (**E**) Schematic of the low-energy band structure in the K valley and proposed impact of different chemical treatments. Untreated S vacancy and untreated O substitution do not possess an optically accessible subgap state. In the untreated S vacancy case, the transition between the defect states ($X_{D2}$) is prohibited as the subgap state is occupied due to the high Fermi-level. With an oxygen substitution, there is no subgap state. TFSI treatment lowers the Fermi level and may remove an oxygen substitution, which allows the defect-to-defect transition, $X_{D2}$, to occur. With PA+TFSI treatment, the Fermi level is lowered and the subgap state is eliminated.

Fig 2.

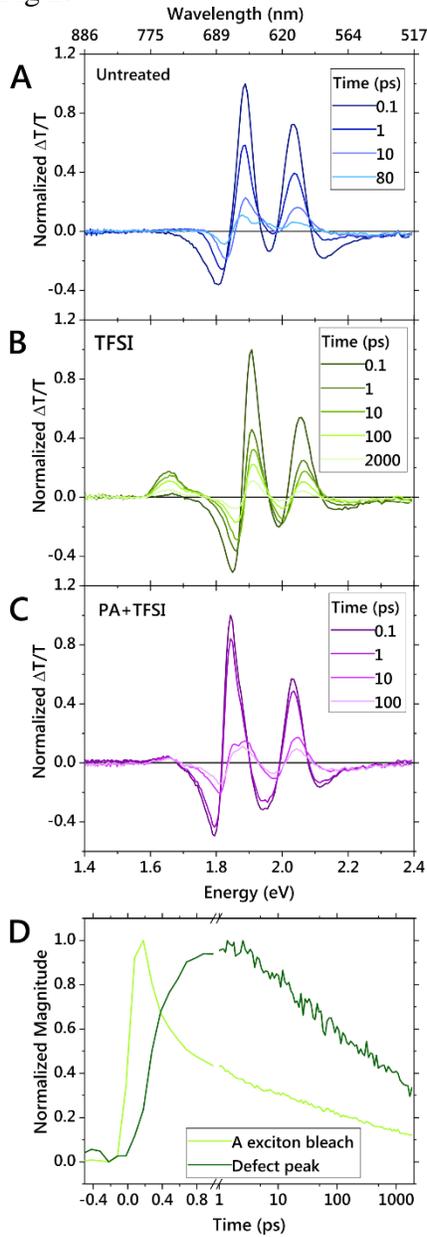

**Figure 2**: **Pump-probe measurements of MoS$_2$ monolayers with different chemical treatments**. Pump-probe spectra of MoS$_2$ untreated (**A**), TFSI-treated (**B**), and PA+TFSI treated (**C**) MoS$_2$. TFSI treatment results in a prominent subgap bleach associated with sulfur vacancy defects. (**D**) Normalized kinetics taken at the A exciton bleach and defect peak in the TFSI-treated sample, illustrating transfer from the band edge to the subgap defect state.

Fig 3:

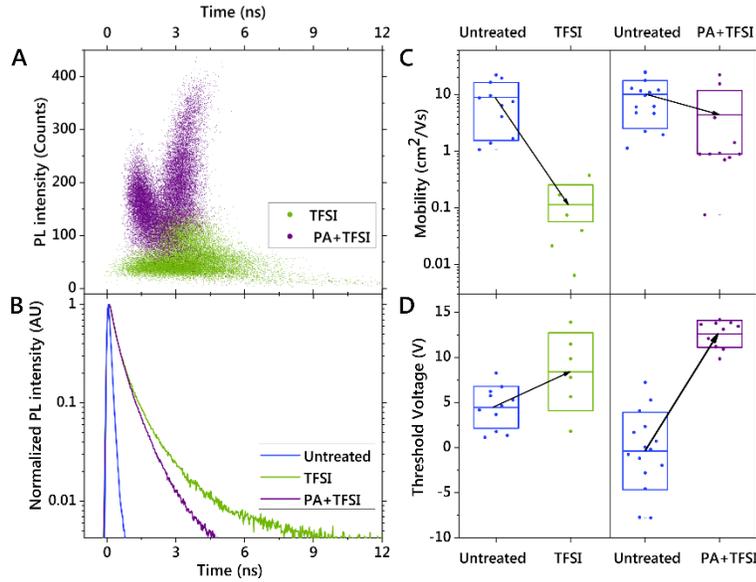

**Figure 3: Time resolved PL and device properties. (A)** Average lifetime versus PL count for TFSI treated and PA+TFSI treated samples. Each data point plots the average lifetime (plotted on the x-axis) and PL intensity (plotted on the y-axis) for a different spot measured from a 2D map taken on monolayer flakes on two different samples for each chemical treatment. **(B)** Example time-correlated single photon counting (TCSPC) trace for untreated, TFSI, and PA+TFSI treated samples. **(C)** Field-effect mobility and **(D)** threshold voltage shift for FET devices before and after treatment, plotting the mean (solid line) and standard deviation (box) of measurements on different devices. Although both TFSI and PA+TFSI treatment decrease the threshold voltage, the PA+TFSI treatment maintains the mobility, whereas TFSI alone results in a significant reduction in mobility.

Fig 4.

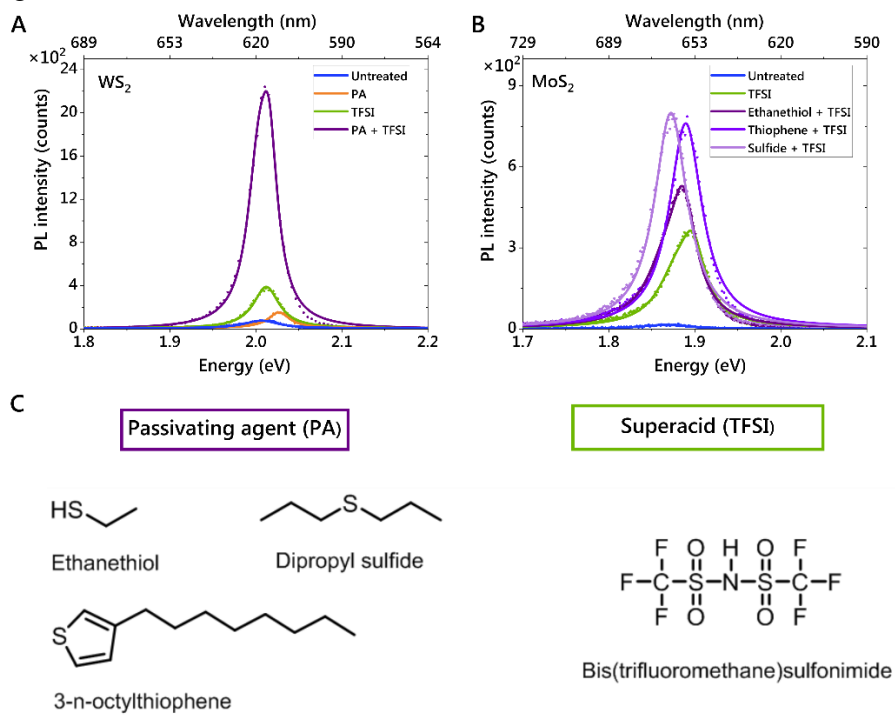

**Figure 4: Generalizability of chemical treatment.** **(A)** PL enhancement of various chemical treatments on $WS_2$. PA alone results in a very minimal increase in PL (also see fig. S5). **(B)** PL enhancement of $MoS_2$ using the different passivating agents outlined in **(C)**.

**Funding:**
We thank the Engineering and Physical Sciences Research Council (EPSRC) and the Winton Programme for the Physics of Sustainability for funding. J.A.- W. acknowledges the support of his Research Fellowship from the Royal Commission for the Exhibition of 1851 and Royal Society Dorothy Hodgkin Research Fellowship. S.D.S acknowledges support from the Royal Society and Tata Group (UF150033). G.D. acknowledges the Royal Society for funding through a Newton International Fellowship. This project has received funding from the European Research Council (ERC) under the European Union's Horizon 2020 research and innovation programme (Grant Agreements 758826 and 756962). Z.L. acknowledges funding from the Swedish research council, Vetenskapsrådet 2018-06610. This work was supported by the Center for Computational Study of Excited State Phenomena in Energy Materials, which is funded by the U.S. Department of Energy,Office of Science, Basic Energy Sciences, Materials Sciences and Engineering Division under Contract No. DE-AC02-05CH11231, as part of the Computational Materials Sciences Program. Work performed at the Molecular Foundry was also supported by the Office of Science, Office of Basic Energy Sciences, of the U.S. Department of Energy under the same contract number. S. R. A. acknowledges Rothschild and Fulbright fellowships. This research used resources of the National Energy Research Scientific Computing Center (NERSC), a DOE Office of Science User Facility supported by the Office of Science of the U.S. Department of Energy under Contract No. DE-AC02-05CH11231.